\documentclass[aps,pra,twocolumn,groupedaddress,showpacs]{revtex4-1}
\usepackage{CJK}
\usepackage{enumerate}
\usepackage{amsmath, amsthm, amssymb}
\usepackage{pifont}
\usepackage{color,calc,graphicx,subfig,epstopdf}

\newcommand{\ket}[1]{| #1 \rangle}
\newcommand{\bra}[1]{\langle #1 |}
\newcommand{\ip}[2]{\langle #1 | #2 \rangle}

\begin{document}

\title{Nonlocality and Entanglement for Symmetric States}

\author{Zizhu Wang}
\email[]{zizhu.wang@telecom-paristech.fr}

\author{Damian Markham}
\email[]{damian.markham@telecom-paristech.fr}
\affiliation{CNRS LTCI, D\'{e}partement Informatique et R\'{e}seaux, Telecom ParisTech, 23 avenue d'Italie, CS 51327,  75214 Paris CEDEX 13, France}

\begin{abstract}
In this paper, building on some recent progress combined with numerical techniques, we shed some new light on how the nonlocality of symmetric states is related to their entanglement properties and potential usefulness in quantum information processing. We use semidefinite programming techniques to devise a device independent classification of three four qubit states into two classes inequivalent under local unitaries and permutation of systems (LUP). We study nonlocal properties when the number of parties grows large for two important classes of symmetric states: the W states and the GHZ states, showing that they behave differently under the inequalities we consider. We also discuss the monogamy arising from the nonlocal correlations of symmetric states. We show that although monogamy in a strict sense is not guaranteed for all symmetric states, strict monogamy is achievable for all Dicke states when the number of parties goes to infinity.
\end{abstract}


\maketitle

\section{Introduction}\label{intro}
Multipartite states are an important resource for many areas of quantum information. Understanding the features that give rise to their usefulness is still a question under much investigation. Entanglement has become recognized as a key feature. However, the question becomes involved in the multipartite settings with different classes of entanglement~\cite{RevModPhys.81.865}, having potentially different roles in recognizing good resources. Intimately related to entanglement is notion of nonlocality~\cite{PhysRev.47.777,bell1964einstein}, though it is known they are not the same~\cite{PhysRevA.40.4277}. Very little is currently known of the richness of multipartite state space and how, if at all, it is exhibited through nonlocality as it is through entanglement theory, with some recent progress in this direction, for example~\cite{PhysRevLett.108.110501,PhysRevLett.108.210407}.

Amongst multipartite states, graph states and symmetric states dominate experimental progress, with experiments with up to 10 qubits~\cite{Lu:2007fk,PhysRevLett.103.020504,PhysRevLett.103.020503,Gao:2010uq}. These two classes represent potentially very different resources for quantum information and have different entanglement features. By exploiting their entanglement properties~\cite{casati2006quantum}, graph states are useful for many tasks such as error correction~\cite{PhysRevA.69.062311}, measurement based quantum computation (MBQC)~\cite{PhysRevLett.86.5188,danos2007measurement,browne2007generalized} and secret sharing~\cite{PhysRevA.78.042309}. In addition to the entanglement properties, nonlocal features of graph states have also been well-studied~\cite{PhysRevLett.95.120405}. Most of these properties are studied via an elegant mathematical tool: the stabilizer formalism~\cite{gottesmanthesis}.

On the other hand, permutation symmetric states occur very often in optics, in the form of Dicke states~\cite{PhysRevLett.103.020504,PhysRevLett.103.020503,PhysRevA.81.032316} and in many-body physics as ground states for example in some Bose-Hubbard models. Similar to graph states, the study of permutation symmetric states can be carried out in an elegant mathematical framework as well: the Majorana representation~\cite{majorana1932atomi}, where symmetric states of $n$ qubits can be represented as $n$ points on the surface of a sphere. In terms of multiparty entanglement properties, much has been learnt using this representation ~\cite{PhysRevLett.103.070503,PhysRevA.81.052315,aulbach2011symmetric,springerlink,PhysRevA.83.042332,PhysRevLett.106.180502,martinthesis}. The nonlocality of symmetric states has been studied recently also by using the Majorana representation~\cite{PhysRevLett.108.210407}, where it was shown that all symmetric states can violate a Bell inequality (it has very recently been shown that all entangled pure states can violate the same inequality~\cite{PhysRevLett.109.120402}). It was also shown in~\cite{PhysRevLett.108.210407} that the degeneracy of the points in the Majorana representation (i.e. when points sit on top of each other) gives persistency of correlations to sets of subsystems. Since degeneracy of Majorana points also separates entanglement classes~\cite{PhysRevLett.103.070503}, this indicates a connection between entanglement classes and nonlocal properties.

As well as the general interest in exploring the texture of multipartite state space, there is some practical interest in understanding the relationship between entanglement and nonlocality. Using the entanglement or nonlocal properties of multipartite states in the real world poses many experimental challenges. Unavoidable experimental inaccuracies like misalignment,  noise and detector inefficiencies can render the outcome of an experiment meaningless. In quantum cryptography, for example, the presence of noise and detector inefficiencies can mask effective effective attacks on the security of the key distribution protocol~\cite{q2007journal,1367-2630-12-11-113026,Lydersen2010fk}. In entanglement theory, misalignment when trying to witness entanglement can lead to mistaken claims of the existence of entanglement~\cite{PhysRevLett.106.250404}. One solution to these problems is to make tangible claims without any assumptions about the measurement device, hence the name \emph{device independent}. There is a natural connection to discussions of nonlocality since Bell type arguments do not rely on any statements about measurements, only their statistics. Using these ideas, device independent proofs and tests have already been used extensively in quantum cryptography and secure communications~\cite{PhysRevLett.98.230501,1367-2630-11-4-045021,PhysRevLett.105.230501,1751-8121-44-9-095305,masanes2011secure,Pironio2010fk}, and device independent entanglement witnesses~\cite{PhysRevLett.106.250404} have been proposed. Recent results have shown device independent tests which are able to discriminate states that are inequivalent under local unitaries and permutation of systems (LUP)~\cite{PhysRevLett.108.110501}.

In this work we further the study of nonlocality of symmetric states using the inequalities and techniques raised in~\cite{PhysRevLett.108.210407}, to study deeper how the nonlocality exposed is related to entanglement classes and the usefulness of the states. We will offer new evidence of connection between the nonlocal properties of states and their entanglement properties through device independent classification of states via violation of Bell inequalities presented in ~\cite{PhysRevLett.108.210407}.  We then look at how the violation of inequalities scales with the number of systems. We see that violation is upper bounded by entanglement because of the form of the inequality, in particular that it has only one positive term. Related to this  we also look at what can be said about the monogamy of the correlations that can be witnessed (a useful property for quantum cryptography~\cite{PhysRevLett.97.170409,PhysRevLett.108.100401}).We see here again that the inequalities of~\cite{PhysRevLett.108.210407} are not suited for showing strict monogamy. Motivated by this we then introduce new inequalities with more positive terms which show good scaling of violation with $n$, implying monogamy in the high $n$ limit.

This paper is organized as follows. In section~\ref{bg}, we give a brief introduction to some concepts and results which we will use in later sections to make the paper self-contained. We recall the main results of~\cite{PhysRevLett.108.210407}: the inequalities to show the nonlocality of all symmetric states and the procedure to find measurement bases to violate them for almost all symmetric states.  Subsequently, in section~\ref{sdp}, we introduce the semidefinite programming (SDP) techniques we use to obtain numerical results, with an example showing the violations of the inequalities introduced in section~\ref{bg} for a class of states. Then in section~\ref{mobius} we use SDP to show how these inequalities allow us to have a device independent discrimination of multipartite entangled state classes, and we see further evidence that degeneracy of Majorana points leads to natural classification with respect to nonlocality. Section~\ref{analytical} shows how the violations of these inequalities scale in the case of large $n$ for two most common sets of symmetric states: the W states and the GHZ states. In section~\ref{monogamy}, we discuss monogamy of entanglement and monogamy of correlations - that is, how much entanglement and correlations can be shared. We see how the inequalities in~\cite{PhysRevLett.108.210407} are not suited to showing strict monogamy, though bounds on how much correlations can be shared can be derived using the methods of~\cite{PhysRevLett.102.030403}. We then generalize a recently presented inequality for W states~\cite{1367-2630-13-5-053054} to all Dicke states and show violation limits to maximal for large $n$ and the correlations from these nonlocal tests are strictly monogamous when the number of parties goes to infinity. We finish with discussions and conclusion.

\section{Background}\label{bg}
We start by giving some background, introducing notation, presenting the inequalities in~\cite{PhysRevLett.108.210407} and how to find the measurement settings to violate them. We consider two dichotomic measurement settings per party, and we use $0$ and $1$ to label both the settings and the outcomes. For measurement settings $M_1, ..M_n$ we denote the probability of getting results $m_1, ...m_n$ as $P(m_1, ...m_n|M_1,...,M_n)$. For example, $P(000|111)$ gives the probability that all three parties obtain result $0$ having measured in setting $1$. The inequality, which we call $\mathcal{P}^n$, is given by
\begin{align}
\mathcal{P}^n:= &P(00\ldots00|00\ldots00)\nonumber\\
-&P(00\ldots00|00\ldots01) \nonumber\\
&\vdots\nonumber\\
-&P(00\ldots00|10\ldots00)\nonumber\\
-&P(11\ldots11|11\ldots11)\leq0,
\end{align}
which must be satisfied by all local hidden variable (LHV) theories. This can be seen since all LHV distributions can be considered as probabilistic mixtures of deterministic local strategies - i.e. ones where $P(m_1, ...m_n|M_1,...,M_n)=\prod_{i}P(m_i|M_i)$ with $P(m_i|M_i)=0$ or $P(m_i|M_i)=1$ - so it is enough to consider these alone~\cite{PhysRevA.64.032112}. Since there is only one positive term in $\mathcal{P}^n$, to have anything greater than zero requires this term to be one. It can easily be seen that this implies that at least one of the negative terms is also one, which gives the desired bound.

Following~\cite{PhysRevLett.108.210407}, to show the violation of $\mathcal{P}^n$ for almost all symmetric states, first we note that all symmetric states of $n$ parties can be written as the sum of permutations of $n$ individual qubits in the Majorana representation ~\cite{majorana1932atomi,RevModPhys.17.237}:
\begin{align}
\ket{\psi}=K\sum_{perm}\ket{\eta_1\ldots\eta_n}\label{decomp},
\end{align}
with the Bloch sphere representation of each qubit state $\ket{\eta_i}$ called a \emph{Majorana point} (MP). We use the notation $\ket{\eta_i^{\perp}}$ to indicate the orthogonal state corresponding to the antipodal point of $\ket{\eta_i}$, such that
\begin{align}
\ip{\eta_i^{\perp}}{\eta_i}=0.
\end{align}
The MPs are then used to find a suitable basis measurements in which show a violation of $\mathcal{P}^n$.

In the prescription described in~\cite{PhysRevLett.108.210407}, for each party, the setting $1$ is chosen to be the basis defined by one of the MPs, say $\ket{\eta_i}$ (associated to outcome $0$) and its orthogonal state $\ket{\eta_i^{\perp}}$ (associated to outcome $1$). Then it is easy to see that
\begin{align}
P(11\ldots11|11\ldots11)=|(\bra{\eta_i^{\perp}})^{\otimes n}\ket{\psi}|^2=0.
\end{align}

To find the basis corresponding to setting $0$, we first notice that the $n-1$ party state
\begin{align}
\ket{\psi'}=\ip{\eta_i}{\psi}
\end{align}
is also a symmetric state, so we can use the same idea. Denoting the $0$ outcome of the basis $0$ by $|0\rangle$, we have
\begin{align}
P(00\ldots00|00\ldots01)&=|(\bra{0})^{\otimes {n-1}}\ip{\eta_i}{\psi}|^2 \nonumber \\
&= |(\bra{0})^{\otimes {n-1}}\ket{\psi'}|^2.
\end{align}
We can then use the Majorana points of $\ket{\psi'}$  for the $0$ basis, as above, to take probabilities $P(00\ldots00|00\ldots01)$ to $P(00\ldots00|10\ldots00)$ zero. Proposition~1 in~\cite{PhysRevLett.108.210407} guarantees that there exists such a choice which also makes $P(00\ldots00|00\ldots00)>0$ for all symmetric states except Dicke states. For Dicke states
\begin{align}
\ket{S(n,k)}=\frac{1}{\sqrt{n \choose k}}(\sum_{perm}\ket{\underbrace{0\ldots0}_{n-k}\underbrace{1\ldots1}_{k}}),
\end{align}
this procedure no longer applies, but $\mathcal{P}^n$ can still be violated by another basis choice (also found in~\cite{PhysRevLett.108.210407}).

In the case that not all $\ket{\eta_i}$ are distinct, we say the state $\ket{\psi}$ is \emph{degenerate}. If $d$ MPs sit on top of each other, we say there is degeneracy $d$. We can extend $\mathcal{P}^n$ to reflect the degeneracy, defining the extended inequality as:
\begin{align} \label{def: degQ}
\mathcal{Q}_{d}^n := &\mathcal{P}^n - P(\underbrace{11\ldots1}_{n-1}|\underbrace{11\ldots1}_{n-1}) - ... - P(\underbrace{11\ldots1}_{n-d+1}|\underbrace{11\ldots1}_{n-d+1})\nonumber\\
& \leq 0.
\end{align}

To calculate the new probabilities, we trace out one party for each term (because of the symmetry of the state, it does not matter which party we trace out):
\begin{align}
P(\underbrace{11\ldots1}_{n-1}|\underbrace{11\ldots1}_{n-1})&=Tr(\rho_1 \underbrace{M_2^{00}\otimes\ldots\otimes M_n^{00}}_{n-1})\nonumber\\
&\vdots\nonumber\\
P(\underbrace{11\ldots1}_{n-d+1}|\underbrace{11\ldots1}_{n-d+1})&=Tr(\rho_{d-1} \underbrace{M_d^{00}\otimes\ldots\otimes M_n^{00}}_{n-d+1}),
\end{align}
where $\rho_1=Tr_i(\ket{\psi}\bra{\psi})$, $\rho_2=Tr_j(\rho_1)$, etc. It can easily be seen that if the MP chosen for basis $1$ has degeneracy $d$, then these probabilities will be zero, hence, the same measurements will lead to a violation of $\mathcal{Q}_{d}^n$. In this sense the persistency of the correlations to subsystems is guaranteed by the degeneracy of MPs.

The connection with degeneracy makes a connection to entanglement classes. Since the degeneracy of points is something that cannot change under SLOCC, states with different degeneracy belong to different classes~\cite{PhysRevLett.103.070503}. As noted in~\cite{PhysRevLett.108.210407}, the above discussion has an interpretation that having at least one MP with degeneracy $d$ automatically means it is possible to violate $\mathcal{Q}_{d}^n$, giving some kind of operational significance to the class. We will study this further in sections III and IV to understand this more.

We will also be interested in what the maximum violation of these inequalities can be. With respect to this question there is a straightforward, though perhaps surprising, bound given be the geometric measure of entanglement~\cite{PhysRevA.80.032324}
\begin{eqnarray} \label{Eqn: Def Eg}
E_g(|\psi\rangle) = \min_{|\Phi\rangle \in {\rm Pro}} -\log_2 (|\langle \Phi | \psi \rangle |^2),
\end{eqnarray}
where ${\rm Pro}$ is the set of product states. It is apparent from the definitions above that for a state $\ket{\psi}$ with entanglement $E_g$, the violations of both $\mathcal{P}^n$ and $\mathcal{Q}_{d}^n$ are bounded by $P(00\ldots00|00\ldots00)$, which in turn is bounded by $\frac{1}{2^{E_g}}$. That is,
\begin{align}
&\mathcal{P}^n\leq \frac{1}{2^{E_g}}&\mathcal{Q}_d^n\leq \frac{1}{2^{E_g}}
\end{align} 
Thus, states with very high entanglement necessarily violate at best by a small amount. The geometric measure is easy to calculate for symmetric states~\cite{PhysRevA.80.032324,aulbach2010maximally,springerlink,PhysRevA.83.042332,martinthesis}, with their entanglement properties with respect to the geometric measure relatively well-known~\cite{PhysRevA.83.042332,springerlink,martinthesis}. However, as we will see in later sections (Figures~\ref{000t_plot} and~\ref{ghz_w_plot}), the geometric measure does not always give a good bound on the violations of $\mathcal{P}^n$ and $\mathcal{Q}_{d}^n$, because of the presence of many negative terms in their expressions. In particular this is true with scaling in number of parties, $n$, where we see that violation decreases as entanglement increases in section~\ref{analytical}. This will later have implications on what can be said about monogamy in section~\ref{monogamy}. The bound from the entanglement will later motivate the introduction of new inequalities with more positive terms so that large violation is possible.

\section{Semidefinite programming techniques}\label{sdp}
In this section we introduce the techniques used to find numerical bounds on the violation of $\mathcal{P}^n$ and $\mathcal{Q}_{d}^n$ for given states. These will, in turn, be used to look at trends in violation with degeneracy, and also to show device independent separation of state classes.

Semidefinite programming was developed in the 90s as a tool to study convex optimization problems~\cite{vandenberghe1996semidefinite}. The method has been adapted in early 2000s as a way to numerically find the global extrema of a real-valued polynomial~\cite{lasserre2001global}. Also around this time, the study of multiparty nonlocality produced increasingly complex results, which made it hard to obtain analytical properties about various multiparty Bell inequalities. As a result, numerical studies about the optimality and violations of these inequalities began to emerge~\cite{PhysRevA.64.032112}~\cite{PhysRevA.64.014102}. In 2006, Wehner~\cite{PhysRevA.73.022110} used SDP as an analytical tool to both prove the original Tsirelson bound for the CHSH inequality and to find new bounds for the generalized CHSH inequality with $n$ settings and 2 outcomes per setting. Since then, SDP has been employed as a numerical tool to study various aspects of multiparty entanglement and features of multiparty nonlocality, for example in ~\cite{PhysRevLett.98.010401}~\cite{PhysRevLett.100.210503}. A recent paper~\cite{PhysRevLett.108.110501} used SDP to show that one can distinguish two different classes of entangled states based on violations of Bell inequalities.

For our purposes, we employ a similar technique to the one used in~\cite{PhysRevLett.108.110501}. Since, without loss of generality, we only use projective measurements~\cite{PhysRevA.64.032112} and probabilities instead of expectation values, the measurement operator we use is different.
Suppose each player $i$ can measure either one of two bases and obtain either one of two possible outcomes. We model these four different situations by four measurement operators:
\begin{align}
M_i^{00}&=\frac{1}{2}(\mathbb{I}_i+\alpha_{i0}\mathcal{X}_i+\beta_{i0}\mathcal{Y}_i+\gamma_{i0}\mathcal{Z}_i)\nonumber\\
M_i^{01}&=\frac{1}{2}(\mathbb{I}_i-\alpha_{i0}\mathcal{X}_i-\beta_{i0}\mathcal{Y}_i-\gamma_{i0}\mathcal{Z}_i)\nonumber\\
M_i^{10}&=\frac{1}{2}(\mathbb{I}_i+\alpha_{i1}\mathcal{X}_i+\beta_{i1}\mathcal{Y}_i+\gamma_{i1}\mathcal{Z}_i)\nonumber\\
M_i^{11}&=\frac{1}{2}(\mathbb{I}_i-\alpha_{i1}\mathcal{X}_i-\beta_{i1}\mathcal{Y}_i-\gamma_{i1}\mathcal{Z}_i),
\end{align}
where $M_i^{jk}$ denotes the player $i$ chooses to measure in basis $j$ and obtains the outcome $k$, and $v_i^0=(\alpha_{i0},\beta_{i0},\gamma_{i0})$, $v_i^1=(\alpha_{i1},\beta_{i1},\gamma_{i1})$ are two unit vectors in $\mathbb{R}^3$.

Now we can write the probabilities in $\mathcal{P}_n$ using these single-qubit measurement operators:
\begin{align}
P(0\ldots0|0\ldots0)&=Tr(\rho M_1^{00}\otimes\ldots\otimes M_n^{00})\nonumber\\
P(0\ldots0|0\ldots1)&=Tr(\rho M_1^{00}\otimes\ldots\otimes M_n^{10})\nonumber\\
\vdots\nonumber\\
P(0\ldots0|1\ldots0)&=Tr(\rho M_1^{10}\otimes\ldots\otimes M_n^{00})\nonumber\\
P(1\ldots1|1\ldots1)&=Tr(\rho M_1^{11}\otimes\ldots\otimes M_n^{11}),
\end{align}
where $\rho=\ket{\psi}\bra{\psi}$ is the density matrix of a $n$-qubit permutation symmetric state $\ket{\psi}$. Rewriting $\mathcal{P}_n$ this way results in a vector polynomial of $2n$ variables ($v_i^0$ and $v_i^1$ for each $i$) \begin{align}
&\mathcal{V}(v_1^0,v_1^1,\ldots, v_n^0, v_n^1)=\nonumber\\
&Tr(\rho M_1^{00}\otimes\ldots\otimes M_n^{00})\nonumber\\
-&Tr(\rho M_1^{00}\otimes\ldots\otimes M_n^{10})\nonumber\\
\vdots\nonumber\\
-&Tr(\rho M_1^{10}\otimes\ldots\otimes M_n^{00})\nonumber\\
-&Tr(\rho M_1^{11}\otimes\ldots\otimes M_n^{11}).
\end{align}
The goal of an SDP program is to maximize $\mathcal{V}(v_1^0,v_1^1,\ldots, v_n^0, v_n^1)$, subject to the constraint that the Gram matrix formed by the vectors $v_i^0$ and $v_i^1$ is positive semidefinite~\cite{horn1990matrix}.

\begin{figure}[htbp]
\begin{center}
\includegraphics[width=230px,keepaspectratio=true]{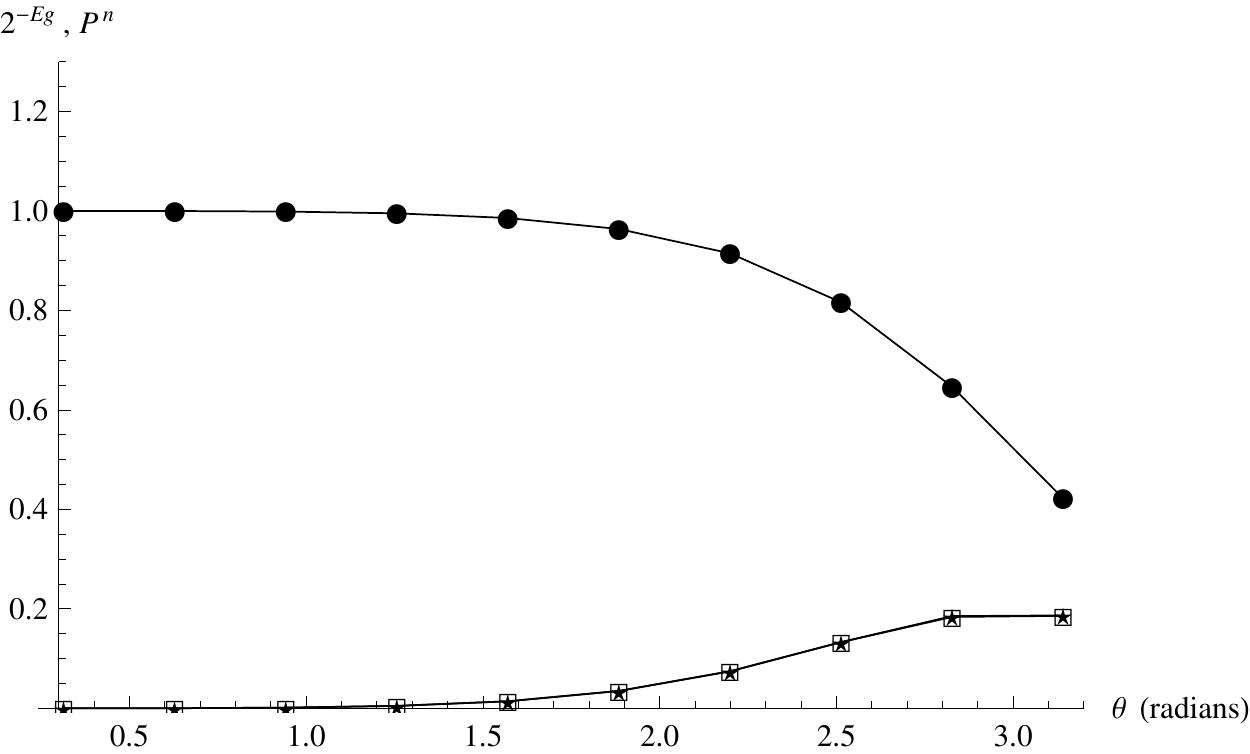}
\caption{A comparison of $\frac{1}{2^{Eg}}$ (\ding{108}), the violation of $\mathcal{P}^4$ (\ding{72}) and $\mathcal{Q}^4_3$ (\ding{113}) for the states $\ket{000\theta}$, when $\theta$ varies from $0$ to $\pi$. }
\label{000t_plot}
\end{center}
\end{figure}

As a first example of the use of SDP, we compute the violation of $\mathcal{P}^4$ and $\mathcal{Q}^4_3$ by a special set of states: the states $\ket{000\theta}=K\sum_{perm}\ket{0}\otimes\ket{0}\otimes\ket{0}\otimes(cos(\frac{\theta}{2})\ket{0}+sin(\frac{\theta}{2})\ket{1})$, with three MPs at the north pole and the other MP varies from $\ket{0}$ (a product state) to $\ket{1}$ (the 4-party W state). For this set of states, the geometric measure of entanglement is easy to calculate by simply searching product symmetric states~\cite{PhysRevA.80.032324}. It should be noted, however, the high level of degeneracy of the state $\ket{000\theta}$ makes it difficult for the SDP program to compute a good bound. This is probably due to the fact that the polynomial defining the SDP problem has high degeneracy, carried over from the degeneracy of the state. The SDP solver we use, SDPNAL, is known to be inaccurate when the optimal solutions are degenerate~\cite{zhao2010newton}. To ease computation for this example, we assume that every player measures in the same bases, which numerically seems to be a reasonable assumption. When we insist every player measures in the same bases, most of the values computed by SDP and plotted in Fig.~\ref{000t_plot} can be certified numerically, meaning there are quantum measurements which can achieve these values. The results are shown in Fig.~\ref{000t_plot}. Note that in later sections we will not make this assumption (unless stated explicitly), here we do so simply as an example to see trends.

It can be seen that the violations of $\mathcal{Q}^4_3$ follow very closely the violations of $\mathcal{P}^4$. Note that in the earlier discussion of degeneracy, where we argued that degeneracy guarantees violation of $\mathcal{Q}^n_d$ (which we can understand as correlations persisting to fewer numbers of parties) we were looking at a particular prescription of measurements, which may not be maximal. In these numerics we have searched over all bases (assuming players all measure in the same basis), indicating that the maximum violation of $\mathcal{Q}^n_d$ is also persistent in correlations of subsets of parties. We also notice that the upper bound given by entanglement is closer to the violation as the angle tends towards $\pi$.

\section{Device independent classification of states}\label{mobius}

Although there are clear connections between the violation of $Q^n_d$ and SLOCC entanglement classes through degeneracy of MPs - degeneracy $d$ guarentees always violation of  $Q^n_d$ - the relationship is not as clear as we might like. An immediate question is the one raised above, the violation of $Q^n_d$ is guaranteed by the prescription using the Majorana representation, but what about the maximal violation? Can we say that degeneracy guarantees that the level of violation stays high? Although we no longer have the analytic tools for general violation, we will see that numerics seems to indicate this is the case, at least for W states. A deeper question though is what we can really understand from this. We would really like to know if it is possible to use these ideas and results to separate classes of states - so that different classes can really be differentiated by their nonlocal properties. This would lead to new ways of searching for new applications of states, as well as ways of probing the texture of multipartite states. To answer this, we will first go more into the subtle questions surrounding the classification of states, and then we will see some examples of how some separation of classes can be made.

On a practical level, it seems clear that different multipartite entangled states have different entanglement and locality properties. Famously GHZ states are highly nonlocal, but are highly sensitive to loss of systems - losing even one system takes them to a separable (hence `local' state), whereas W states do not have the same extreme nonlocality~\cite{PhysRevA.63.022104}, but losing systems does not destroy the entanglement. In turn, different types of states may have different uses for quantum information.

The question of how to classify states in terms of entanglement and locality is a difficult one, particularly when we want to talk about how different `classes' might be meaningful either for different quantum information tasks, or their potential roles in many-body physics. Within entanglement theory, the most standard approach is to define two states as equivalent if they can be mapped to one another using only local operations and classical communication (LOCC), with some non-zero probability. This method of classification leads to what are called SLOCC classes of states (the S standing for Stochasitic)~\cite{PhysRevA.62.062314}. Intuitively this classification is appealing since it separates states which cannot be reached from each other in the distributed setting, even with the aid of classical communication.

In terms of how one might classify states with respect to locality, there are several approaches. The standard setting for locality questions is one in which parties are not allowed to communicate classically - at least not after they have been told what bases to measure in, they may do before hand, for example to share classical randomness. Several options arise. In~\cite{PhysRevA.83.022328} it is proposed that a reasonable classification is to consider equivalence under local unitaries and permutation of systems (we denote this LUP). One may also consider states equivalent under local operations, which is in turn equivalent to local unitaries (we donate this LU). When considering correlations alone, without necessarily taking recourse to quantum states, in~\cite{PhysRevLett.109.070401} a classification is presented called wiring and classical communication prior to inputs (WCCPI) - the wiring is essentially the idea of using multiple copies of the resource (which could be a quantum state or `box' giving a certain probability distribution) and allowing different ways of combining them. We do not consider the WCCPI classification further here, and rather focus on single copy classifications.

For all the classifications mentioned above, however, several difficulties emerge, which seem to limit their usefulness. First of all, there can be an infinite continuum of classes (for LUP and LU this is already true for two quibits, for SLOCC it is true for four or more~\cite{PhysRevA.62.062314}). Second, and related to this, it is possible to have two states which are arbitrarily close to each other which are in different classes. This means that two states, which behave in almost exactly the same way for all possible experiments, can be in different classes. It is clear then that it is not possible to separate all classes of states in terms of their physical properties and in turn that the physical properties cannot be sensitive to all these classifications. Nevertheless, there does seem to be some difference between states, which can be identified through these classifications.
For example, as we saw earlier, states of certain classes guarantee resistance of correlations to loss of systems, for both the LUP~\cite{PhysRevLett.108.110501} and the SLOCC~\cite{PhysRevLett.108.210407} classifications (through the degeneracy of MPs as mentioned earlier). In~\cite{PhysRevLett.108.110501} this was used to separate two LUP classes in a device independent way.

Here we will use our inequalities to identify different sets of LUP classes of states of four qubits, hence also, in a device independent way. The LUP and LU classifications are well suited to discriminate via inequality violation because the maximum violation of an inequality is searched for over all measurement bases - which is equivalent to searching over all local unitaries. Thus, if we can say that a particular state cannot violate an inequality more than a certain amount (using SDP techniques for example, as we do here), this means that no state in the same LU class can either. If the state is symmetric it also means no state in the same LUP state can either. The states we choose are also in different SLOCC classes (note, however, that the fact that no LU or LUP equivalent state can violate more than the amount we state does not necessarily mean that there does not exist an SLOCC equivalent state which can). Since this is done via violation of Bell-like inequalities - which makes no recourse to what measurements are made, this classification is done in a device independent way.

\begin{figure}[ht]
  \centering
  \subfloat[]{\label{ex:t}\includegraphics[width=120px,keepaspectratio=true]{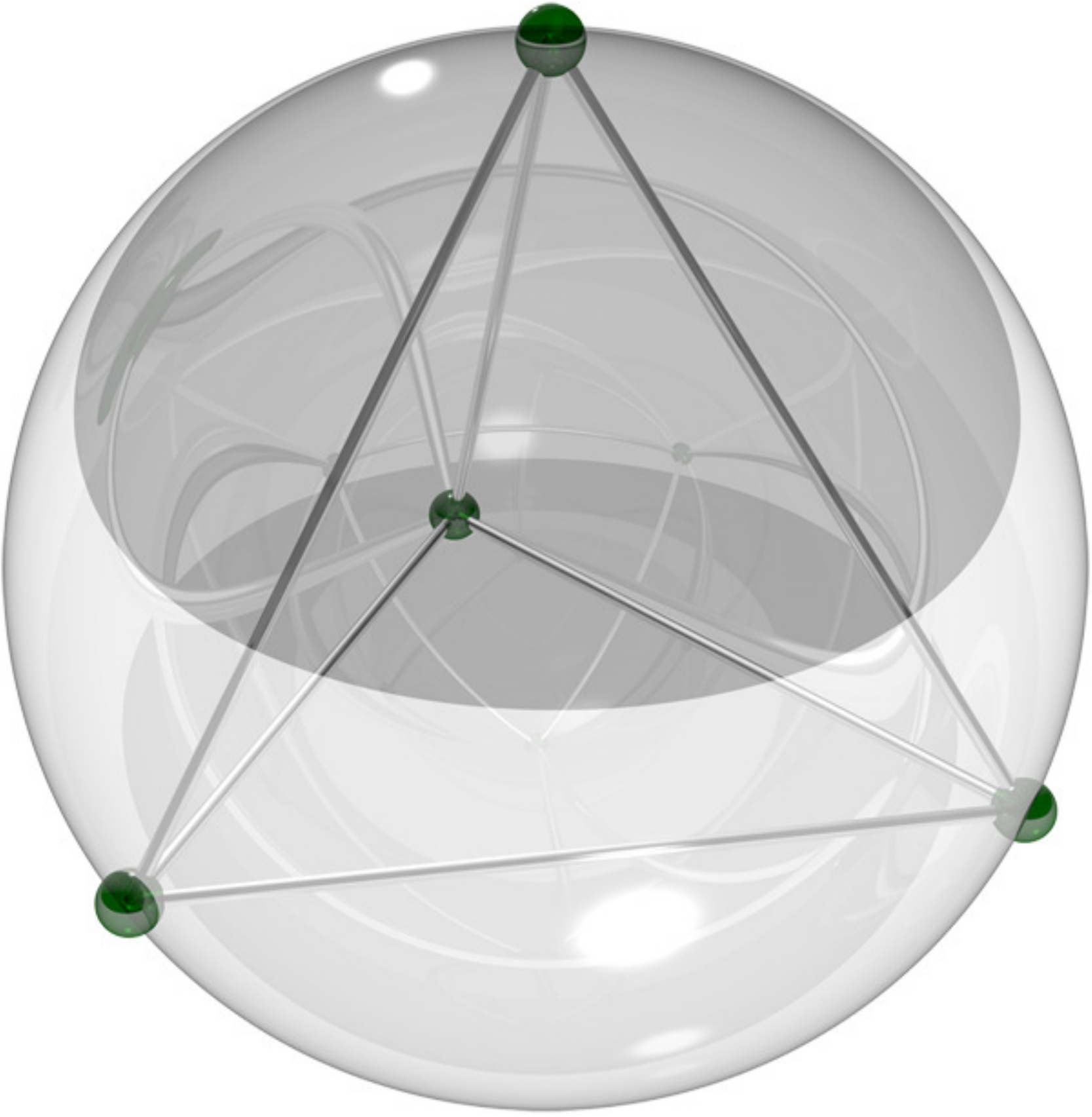}}
  \,
  \subfloat[]{\label{ex:ghz}\includegraphics[width=120px,keepaspectratio=true]{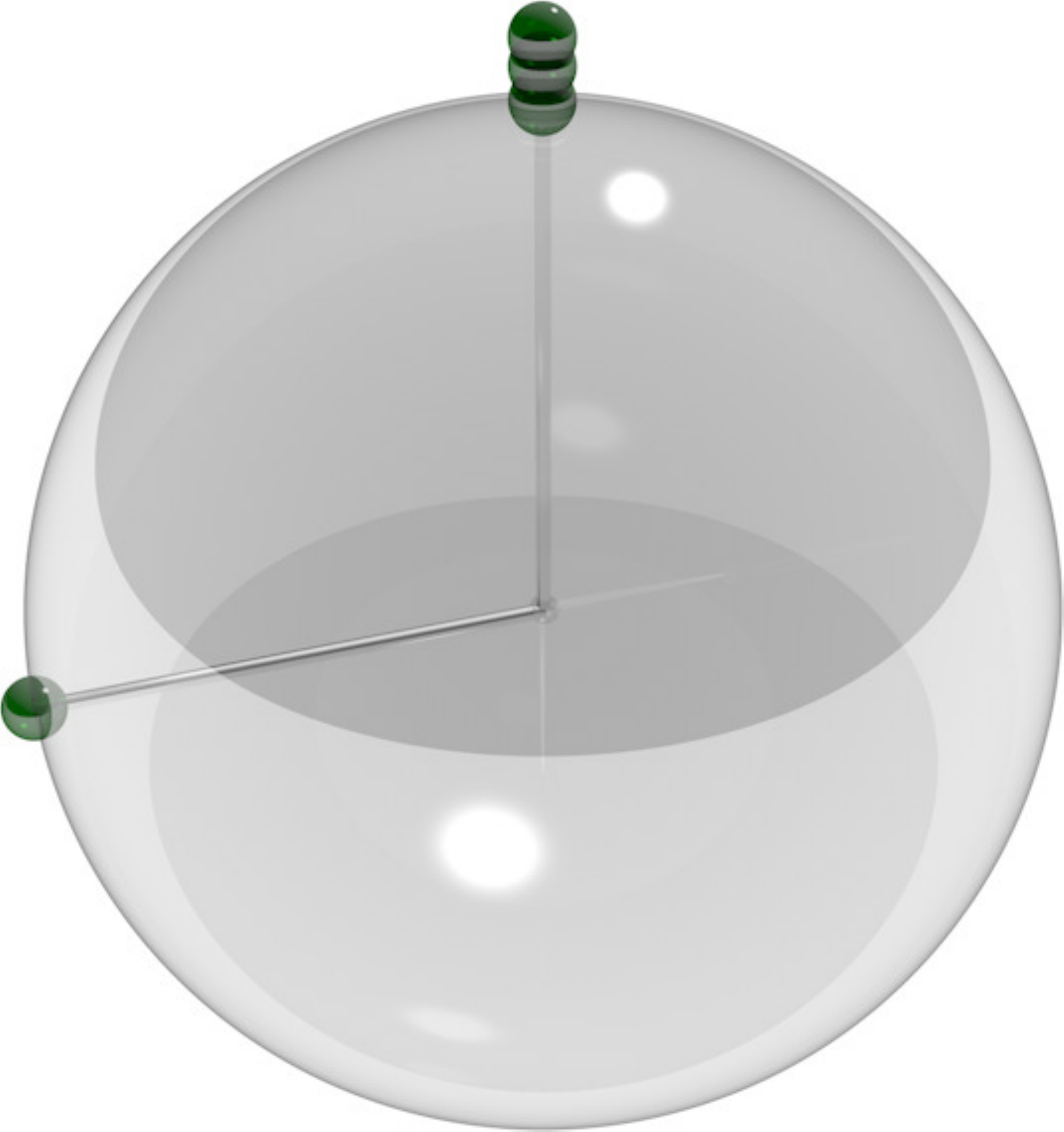}}
  \caption{The tetrahedron state (a) and the state $\ket{000+}$ (b) in the Majorana representation.}
  \label{t_000+}
\end{figure}

\begin{figure}[ht]
  \centering
  \subfloat[]{\label{ex:t}\includegraphics[width=120px,keepaspectratio=true]{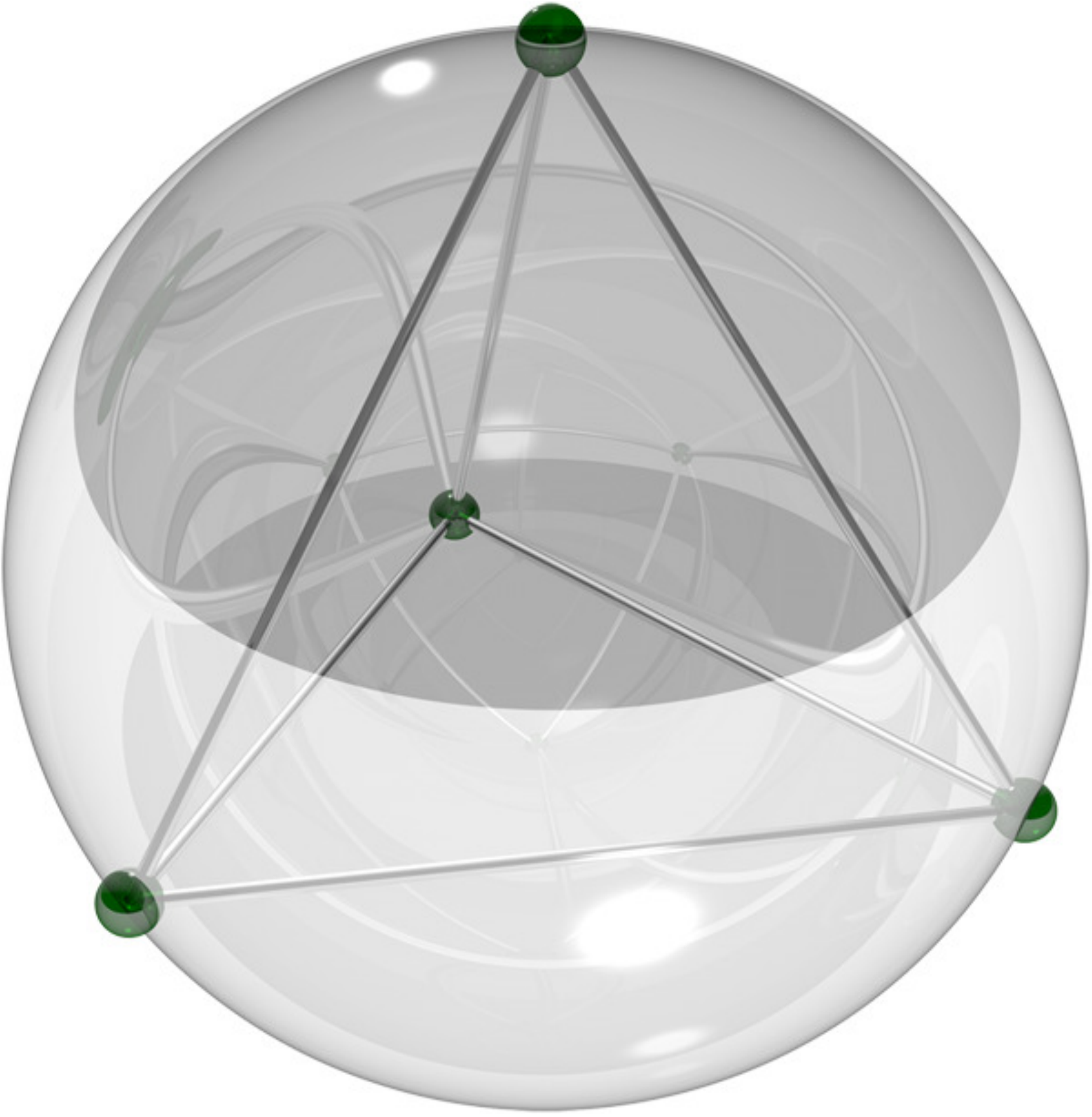}}
  \,
  \subfloat[]{\label{ex:ghz}\includegraphics[width=120px,keepaspectratio=true]{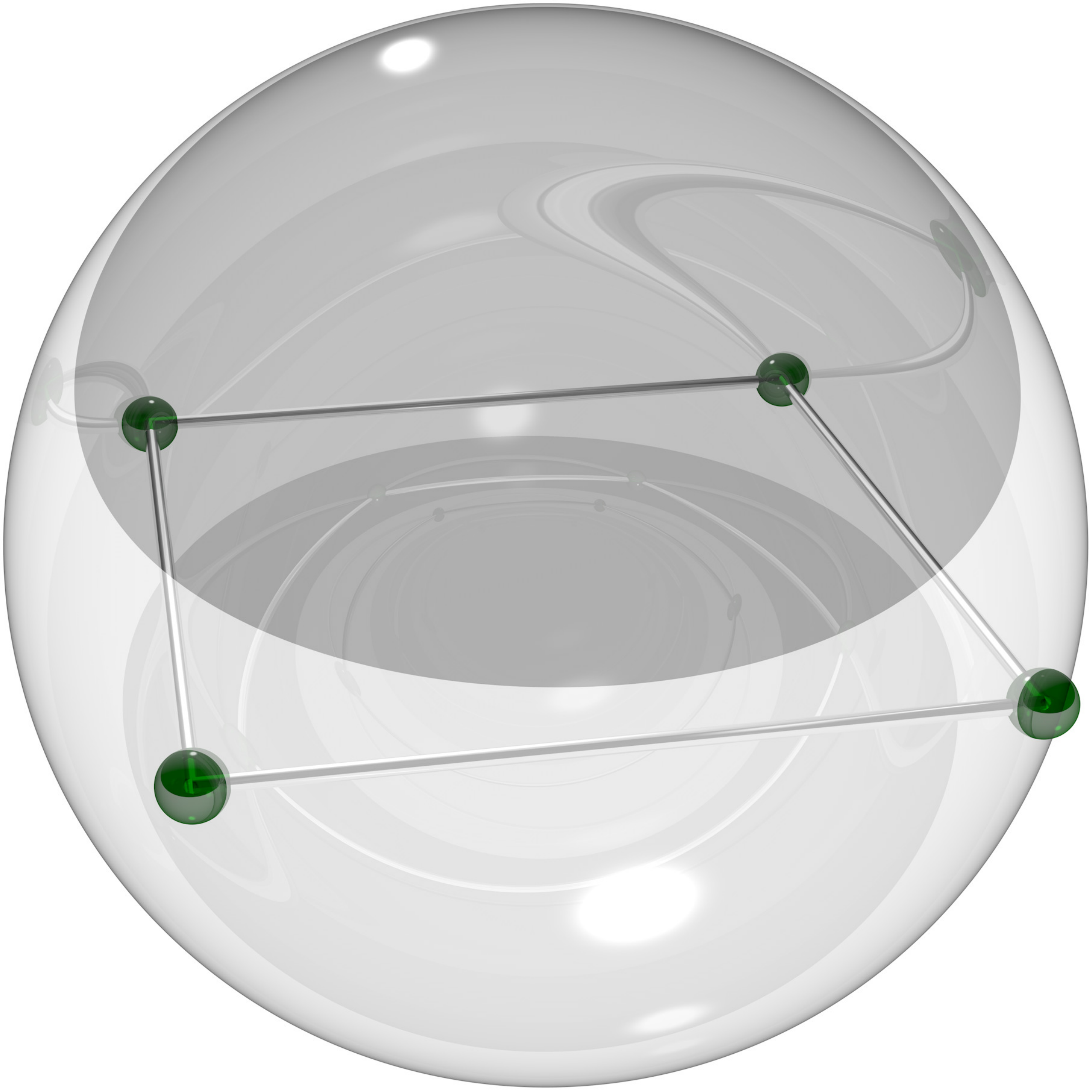}}
  \caption{The tetrahedron state (a) and the 4-qubit GHZ state (b) in the Majorana representation.}
  \label{ghz_t}
\end{figure}

For the classification, we will consider three states: the tetrahedron state $\ket{T}=\sqrt{\frac{1}{3}}\ket{S(4,0)}+\sqrt{\frac{2}{3}}\ket{S(4,3)}$, the 4-qubit GHZ state $\ket{GHZ_4}=\frac{1}{\sqrt{2}}(\ket{0000}+\ket{1111})$, and the state $\ket{000+}=K\sum_{perm}\ket{000+}=\frac{2}{\sqrt{5}}\ket{0000}+\frac{1}{\sqrt{5}}\ket{S(4,1)}$, which are all SLOCC-inequivalent~\cite{PhysRevA.83.042332,PhysRevLett.103.070503,aulbach2011symmetric,PhysRevLett.106.180502}. We will consider them in two groups: one group consists of $\ket{T}$ and $\ket{000+}$, with differing degeneracy, the other group consists of $\ket{T}$ and $\ket{GHZ_4}$, with the same degeneracy. These are represented in Fig.~\ref{t_000+} and \ref{ghz_t} respectively. We will use numerical maximum violation of $\mathcal{P}^4$ and $\mathcal{Q}^4_3$ obtained from SDP to discriminate the states in a device independent way in each group.

The SLOCC-inequivalence of these states can be seen most easily from a recent result~\cite{aulbach2011symmetric}~\cite{PhysRevLett.106.180502}, which has shown that for symmetric states, there is an interesting relationship between SLOCC operations and M\"{o}bius transformations. A \emph{M\"{o}bius transformation} is a function of one variable $z$, defined on the extended complex plane $\mathbb{C}_{\infty}$, which can be written in the form
\begin{align}
f(z)=\frac{az+b}{cz+d},
\end{align}
where $a, b, c ,d$ are complex numbers and $ad-bc\neq0$ (otherwise $f(z)$ is a constant map)~\cite{needhamcomplex}~\cite{beardon2005algebra}. A M\"{o}bius transformation can be seen as a composition of four more elementary steps: translation, complex inversion, expansion and rotation. M\"{o}bius transformations are conformal maps which take circles to circles and preserve the symmetry with respect to circles.
The SLOCC-inequivalence of $\ket{T}$, $\ket{GHZ}$ and $\ket{000+}$ can be seen from the fact that it is not possible to change the degeneracy of MPs via M\"{o}bius transformations (see also~\cite{PhysRevLett.103.070503}).
When we consider $\ket{T}$ and $\ket{GHZ}$, it is clear from Fig.~\ref{ghz_t} that there is no M\"{o}bius transformation connecting their Majorana points: the MPs of $\ket{GHZ_4}$ all lie on the equator, a M\"{o}bius transformation will map them to another circle, but the MPs of $\ket{T}$ clearly do not form a circle.

The equivalence of symmetric states under LU and LUP is given simply by the MP distribution up to rotation of the sphere. This is because any local unitary taking a symmetric state to a symmetric state can be understood as a rotation of the sphere~\cite{PhysRevLett.103.070503,PhysRevA.81.052315} (and that permutation obviously do not change a symmetric state). Thus each of the states we study here are LU and LUP inequivalent. As mentioned, the fact that we search for violation of inequalities over all measurements means that the bounds we present hold for all LU and LUP equivalent states.

For the first group, shown in Fig.~\ref{t_000+}, the results are shown in Table~\ref{t_000+_table}. Note that although we do not restrict the measurement bases for $\ket{T}$, as the degeneracy of the state $\ket{000+}$ is very high, we need to restrict the bases to get realistic SDP bounds.

\begin{table}[htdp]
\begin{center}
\begin{tabular}{|c|c|c|}
\hline
State&$\mathcal{P}^4$&$\mathcal{Q}^4_3$\\
\hline
\hline
$\ket{T}$&0.1745&-0.0609\\
\hline
$\ket{000+}$&0.0142&0.0141\\
\hline
\end{tabular}
\end{center}

\caption{SDP bounds on the maximum violation of $\mathcal{P}^4$ and $\mathcal{Q}^4_3$ for $\ket{T}$ and $\ket{000+}$. Because of computational difficulties the values for $\ket{000+}$ assume that all parties measure in the same basis (numerics indicate this is still optimal). We thus have that a violation of $\mathcal{Q}^4_3$ implies the state is not in the LU class of $|T\rangle$.}
\label{t_000+_table}
\end{table}%

Table.~\ref{ghz_t_table} shows the bounds for $\mathcal{P}^4$ and $\mathcal{Q}^4_3$ for the second group, shown in Fig.~\ref{ghz_t}, obtained using semidefinite programming techniques described in section~\ref{sdp}, without restricting the measurement bases of parties.

\begin{table}[htdp]
\begin{center}
\begin{tabular}{|c|c|c|}
\hline
State&$\mathcal{P}^4$&$\mathcal{Q}^4_3$\\
\hline
\hline
$\ket{T}$&0.1745&-0.0609\\
\hline
$\ket{GHZ_4}$&0.1241&0.0563\\
\hline
\end{tabular}
\end{center}

\caption{SDP bounds on the maximum violation of $\mathcal{P}^4$ and $\mathcal{Q}^4_3$ for $\ket{T}$ and $\ket{GHZ_4}$. We thus have that a violation of $\mathcal{P}^4 > 0.1241$ implies the state is not in the LU class of $|GHZ_4\rangle$, and a violation of $\mathcal{Q}^4_3$ implies the state is not in the LU class of $|T\rangle$.}
\label{ghz_t_table}
\end{table}%

From these tables, one can easily envisage device independent tests to discriminate the LUP classes in each group.

For the first group, because of the restriction on measurement bases, we have a weaker test. Despite our best numerical checks and the seemingly reasonable assumption on the restriction of measurement of bases, we cannot guarantee that if a state has a violation of $\mathcal{P}^4$ greater than $0.0142$, it is not in the LUP $\ket{000+}$ class. However, we can still conclude that if a state violates $\mathcal{Q}^4_3$ then it cannot be in the $\ket{T}$ class, but must be in the $\ket{000+}$ LUP class.

In the second group, if the $\mathcal{P}^4$ test gives a violation $\geq 0.1241$, then the state must not be in the $\ket{GHZ_4}$ LUP class, so must be in the $\ket{T}$ class. Similarly, if the $\mathcal{Q}^4_3$ gives any violation at all, the state cannot be in the $\ket{T}$ LUP class and must be in the $\ket{GHZ_4}$ class. In this case, even though there is no degeneracy, separation can be seen using $\mathcal{Q}^4_3$.

\section{Large $n$ results for $\ket{W_n}$ and $\ket{GHZ_n}$}\label{analytical}

In this section we study the trends of violations of $\mathcal{P}_n$ for W and GHZ states as $n$ gets large. In terms of monogamy and other applications of nonlocal features (for example communication complexity gains~\cite{RevModPhys.82.665}), we are interested in the value of violation - the higher the better. We are interested then to know how violation scales with $n$.

While the use of SDP allows us to study the nonlocality of symmetric states with a few parties, the computational resources required to run the SDP program increase exponentially with the number of parties, which makes it impractical to obtain results for states with more than 4 parties. Luckily, for two commonly studied symmetric states, the W states
\begin{align}
\ket{W_n}=\ket{S(n,1)}=\frac{1}{\sqrt{n}}(\sum_{perm}\ket{\underbrace{0\ldots 0}_{n-1}1}),
\end{align}
and the GHZ states
\begin{align}
\ket{GHZ_n}=\frac{1}{\sqrt{2}}(\ket{\underbrace{0\ldots 0}_{n}}+\ket{\underbrace{1\ldots 1}_{n}}),
\end{align}
it is possible to calculate analytically the violation of $\mathcal{P}_n$ if the measurement bases are those prescribed in section~\ref{bg} and~\cite{PhysRevLett.108.210407}. This allows us to give bounds on the maximum violation possible and see trends. We will use a combination of this and numerics to approximate the best violation.

For the W state, using the bases $\{\ket{+},\ket{-}\}$, $\{\ket{0},\ket{1}\}$ as settings $0$ and $1$ in $\mathcal{P}^n$, we get the violation
\begin{align}
v_w(n)=\frac{n-2}{n\times2^{n}}.
\end{align}

\begin{figure}[htbp]
\begin{center}
\includegraphics[width=230px,keepaspectratio=true]{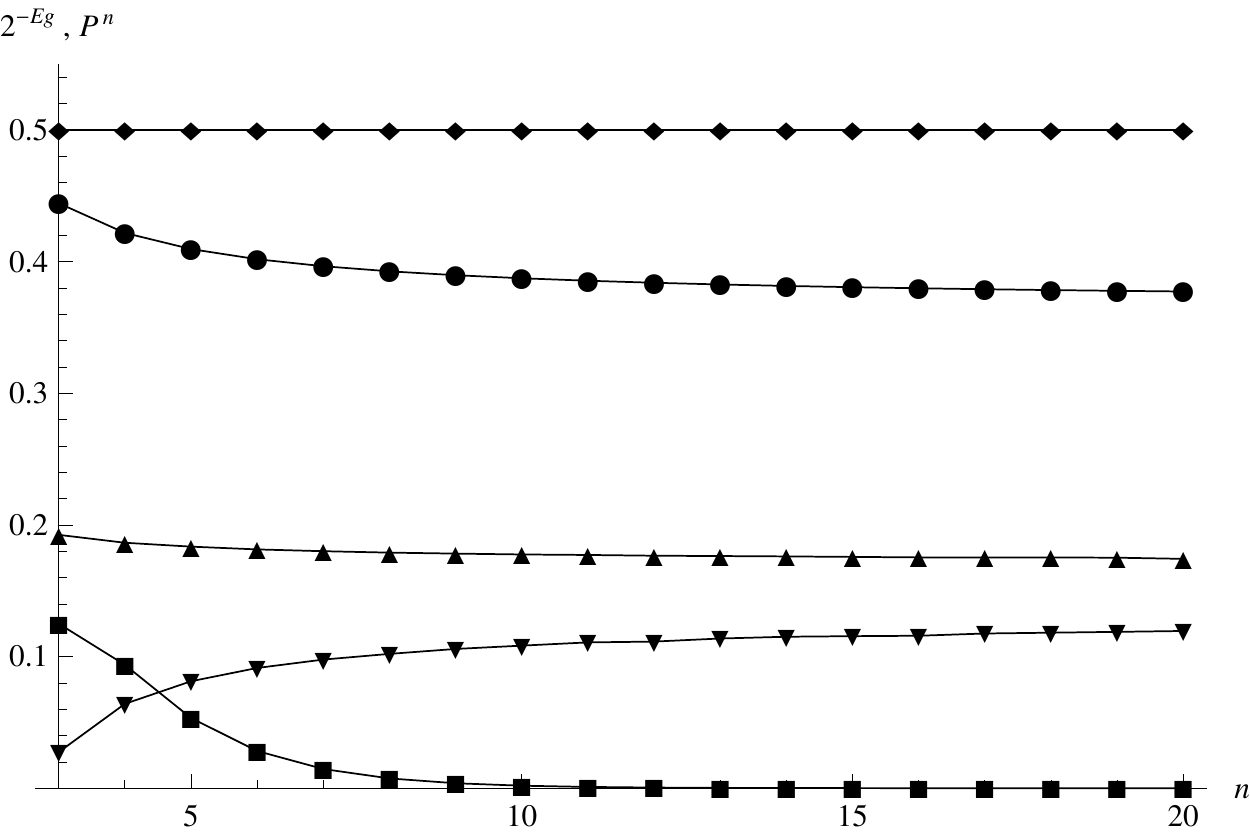}
\caption{Violations of $\mathcal{P}^n$ by the state $\ket{GHZ_n}$ (\ding{110}), with the  numerical violations of $\mathcal{P}^n$ (\ding{115}) and $\mathcal{Q}^n_{n-1}$ (\ding{116}) of $\ket{W_n}$ as a function of $n$ (number of parties), comparing to $\frac{1}{2^{Eg(\ket{W_n})}}$ (\ding{108}) and $\frac{1}{2^{Eg(\ket{GHZ_n})}}$(\ding{117}).}
\label{ghz_w_plot}
\end{center}
\end{figure}

This algebraic violation, while works for all $\ket{W_n}$, is not the optimum violation. By optimizing over the four Euler angles in the two bases, we obtained close to optimal numerical violations of $\mathcal{P}^n$ (\ding{115} in Fig.~\ref{ghz_w_plot}) and $\mathcal{Q}^n_{n-1}$ (\ding{116} in Fig.~\ref{ghz_w_plot}) for W states. It can be seen from the plot that the violations of $\mathcal{P}^n$ is close to the upper bound derived from the geometrical measure of entanglement, $\frac{1}{2^{Eg(\ket{W_n})}}$.

For GHZ states, we can follow the procedure given in section~\ref{bg} to find the bases. Note that the MPs of GHZ states with an even number of parties and an odd number of parties are different. For example, $\ket{+}$ is an MP of $\ket{GHZ_n}$ when $n$ is odd, but not when $n$ is even. Nevertheless, the MPs in both cases are all equally distributed along the equator of the Bloch sphere, allowing us to have a single expression for the bases as a function of $n$. The basis 1, which consists an MP and its antipodal point, is $\{\frac{1}{\sqrt{2}}(\ket{0}-e^{-\mathrm{i}\frac{\pi}{n}}\ket{1}),\frac{1}{\sqrt{2}}(\ket{1}+e^{\mathrm{i}\frac{\pi}{n}}\ket{0})\}$, and the basis 0 is $\{\frac{1}{\sqrt{2}}(\ket{0}-e^{\mathrm{i}\frac{(2n-1)\pi}{n(n-1)}}\ket{1}),\frac{1}{\sqrt{2}}(\ket{1}+e^{-\mathrm{i}\frac{(2n-1)\pi}{n(n-1)}}\ket{0})\}$. Calculating the violation as a function of $n$ (which is just the probability $P(0\ldots0|0\ldots0)$), we have (the \ding{110} line in Fig.~\ref{ghz_w_plot})
\begin{align}
v_g(n)=\frac{1}{2^n}(1+cos(\frac{(2n-1)\pi}{n-1})).
\end{align}
This violation agrees with the best found by numerics.

From Fig.~\ref{ghz_w_plot}, we can see that as $n$ increases, the violations are always well below $\frac{1}{2^{Eg}}$, which follows the trend we noticed in the earlier SDP examples. We also numerically optimized the value of $\mathcal{Q}^n_{n-1}$, which is always negative for GHZ states. This is in stark contrast to the situation for W states, where the violation of $\mathcal{Q}^n_{n-1}$ stays slightly below the violation of $\mathcal{P}^n$. One interpretation of this phenomenon is that $\mathcal{Q}^n_{d}$ is closely related to the degeneracy of the state, and can be used as a `witness' of degeneracy for these states.

\section{Monogamy}\label{monogamy}
\subsection{General Discussions}
In sections~\ref{mobius} and~\ref{analytical}, we studied the nonlocality and entanglement for symmetric states from the perspective of different ``types'' in each context. There is another property, defined for both contexts, that highlights yet another interesting aspect of the relationship between nonlocality and entanglement: that is the concept of monogamy~\cite{PhysRevA.61.052306}~\cite{Toner08012009} (for a review see~\cite{springerlink:10.1007/s11128-009-0161-6}).

As its name suggests, monogamy measures the ``exclusiveness'' of entanglement or correlations, that is, how well they can be shared. For example if two parties share a maximally entangled state or a maximally correlated Popescu-Rohrlich (PR) box~\cite{popescu1994quantum}, the entangled systems or PR box cannot be entangled or correlated to anything else. In recent years it has been recognized as a key ingredient to the usefulness of states for example in security and device independent security scenarios~\cite{PhysRevLett.97.170409,Pironio2010fk,1751-8121-44-9-095305,PhysRevLett.108.100401}. The idea being that if the correlations cannot be shared, that means that the eavesdropper is uncorrelated with the honest parties, so the information they share will not be leaked to the eavesdropper.

Monogamy of entanglement is a property of a particular quantum state. It measures the intra-subgroup entanglement tradeoff with respect to a suitably chosen entanglement measure. The most famous such measure is the tangle $\tau$ introduced in~\cite{PhysRevA.61.052306}, which measures the entanglement across a bipartition. The CKW inequality, proposed in~\cite{PhysRevA.61.052306} as a conjecture and proved recently in~\cite{PhysRevLett.96.220503}, states that for all pure entangled states, the sum of all bipartite tangles between one party $A$ and $n$ parties $\{B_1,\ldots,B_n\}$ is less than or equal to the tangle between $A$ and all $B_i$ considered as a whole:
\begin{align}
\tau(\rho_{AB_1})+\tau(\rho_{AB_2})+\ldots+\tau(\rho_{AB_n})\leq\tau(\rho_{A(B_1\ldots B_n)}).\label{ckw}
\end{align}
Although it is known that symmetric states like the W state can saturate this inequality, not all states which saturate this inequality are symmetric. 

The monogamy of 3-qubit symmetric states have been studied recently~\cite{PhysRevA.85.012103}, using a different measure of quantum correlations, called the quantum deficiency (related to quantum discord~\cite{PhysRevLett.88.017901}).  It was shown that SLOCC equivalent states do not necessarily have the same monogamy relation with respect to this measure. Here we focus on correlations of the measurement results directly (which we call simply ``monogamy of correlations"). 

Monogamy of correlations is normally defined in the context of correlations arising from probability distributions, without explicitly referring to quantum states and measurements. Intuitively, monogamy says that strong correlations cannot be shared. In a strict sense, we say an $n$-partite distribution, $P(a_1,\ldots,a_n|A_1,\ldots,A_n)$, is monogamous~\cite{PhysRevA.71.022101}~\cite{PhysRevLett.108.100401}, if the only nonsignaling extension to $n+1$ parties $P(a_1,\ldots,a_n,a_{n+1}|A_1,\ldots,A_n,A_{n+1})$ is the trivial one, i.e. such that
\begin{align}
&P(a_1,\ldots,a_n,a_{n+1}|A_1,\ldots,A_n,A_{n+1})\nonumber\\
=&P(a_1,\ldots,a_n|A_1,\ldots,A_n)P(a_{n+1}|A_{n+1}).\label{strict_mon}
\end{align}

For all possible measurement settings $A_k$ and $A'_k$ for party $k$, the nonsignaling condition can be stated as
\begin{align}
&P(a_1,\ldots,a_{k-1},a_{k+1}\ldots,a_{n}|A_1,\ldots,A_{k-1},A_{k+1},\ldots,A_{n})\nonumber\\
&=\sum_{a_k}P(a_1,\ldots,a_k,\ldots,a_{n}|A_1,\ldots,A_k,\ldots,A_{n})\nonumber\\
&=\sum_{a_k}P(a_1,\ldots,a_k,\ldots,a_{n}|A_1,\ldots,A'_k,\ldots,A_{n}).\label{no-signal}
\end{align}
That is, when tracing out one system, $k$, to get the marginal distributions, it does not matter which measurement setting $A_k$ is used.

This strict sense of monogamy is guaranteed if an inequality reaches its algebraic maximum~\cite{PhysRevLett.97.170409}. Indeed, this fact is used to show monogamy for several states via several inequalities including GHZ states~\cite{PhysRevLett.97.170409}~\cite{PhysRevLett.108.100401}~\cite{Toner08012009}. However, the inequalities $\mathcal{P}^n$ and $\mathcal{Q}^n_d$ here cannot show strict monogamy in this way, simply because no quantum state can ever achieve the algebraic bound, as the bound is given by the entanglement. In the following subsection we will develop another set of inequalities for which this idea does work.

Even if not demanding strict monogamy of correlations, it is possible to bound how well correlations can be shared.
In~\cite{PhysRevLett.102.030403}, a bound is presented covering general nonsignaling theories by demanding tradeoffs of correlations in a multipartite setting, analogous to the monogamy of multipartite entanglement.
To apply these results to our inequality, we will follow the prescription given in~\cite{PhysRevLett.102.030403}. First we rewrite our inequality to make all terms positive:
\begin{align}
\mathcal{P}^n&=P(0\ldots0|0\ldots0)\nonumber\\
&-(1-\sum_{a_1,\ldots,a_n\neq{0\ldots0}}P(a_1,\ldots,a_n|0\ldots1))\nonumber\\
&\vdots\nonumber\\
&-(1-\sum_{a_1,\ldots,a_n\neq{0\ldots0}}P(a_1,\ldots,a_n|1\ldots0))\nonumber\\
&-(1-\sum_{a_1,\ldots,a_n\neq{1\ldots1}}P(a_1,\ldots,a_n|1\ldots1)).
\end{align}

By keeping all the probabilities on the left hand side and moving everything else to the right hand side, we define the inequality
\begin{align}
\mathcal{P}^{n'}&=P(0\ldots0|0\ldots0)\nonumber\\
&+\sum_{a_1,\ldots,a_n\neq{0\ldots0}}P(a_1,\ldots,a_n|0\ldots1)\nonumber\\
&\vdots\nonumber\\
&+\sum_{a_1,\ldots,a_n\neq{0\ldots0}}P(a_1,\ldots,a_n|1\ldots0)\nonumber\\
&+\sum_{a_1,\ldots,a_n\neq{1\ldots1}}P(a_1,\ldots,a_n|1\ldots1)\nonumber\\
&\leq n+1.
\end{align}

Now we can partition the parties into two groups: group $A$ with $k$ parties and group $B$ with $n-k$ parties. Consider a single group $A$ which is possibly correlated with multiple identical $B^i$. The multiparty monogamy relation of~\cite{PhysRevLett.102.030403} tells us that for any nonsignalling probability distribution for $n>2$
\begin{align}
\sum_{i=1}^{n-k+2}\mathcal{P}^{n'}(A,B^i)\leq (n-k+2) (n+1),\label{broad_mon}
\end{align}
where $i$ runs over the possible combinations of measurement settings of $n-k$ parties that make up each $B^i$.

For $\mathcal{Q}_{d}^n$, we can treat the $d-1$ extra probabilities as marginals of probabilities involving $n$ parties:
\begin{align}
P(\underbrace{1\ldots1}_{n-d+1}|\underbrace{1\ldots1}_{n-d+1})=\sum_{b_1,\ldots,b_{d-1}}P(\underbrace{1\ldots1}_{n-d+1}\underbrace{b_1\ldots b_{d-1}}_{d-1}|\underbrace{1\ldots1}_{n}),
\end{align}
which leads to the inequality for $\mathcal{Q}_{d}^{n'}$:
\begin{align}
\mathcal{Q}_{d}^{n'}&=\mathcal{P}^{n'}+\sum_{a_1,\ldots,a_{n-1}\neq{1\ldots1}}P(a_1,\ldots,a_{n-1},b_1|\underbrace{1\ldots1}_{n})\nonumber\\
&\vdots\nonumber\\
+&\sum_{a_1,\ldots,a_{n-d+1}\neq{1\ldots1}}P(a_1,\ldots,a_{n-d+1},b_1,\ldots,b_{d-1}|\underbrace{1\ldots1}_{n})\nonumber\\
&\leq n+d.
\end{align}

Because the expression for $\mathcal{Q}_{d}^{n'}$ does not increase the number of settings for $B$, we have the monogamy inequality for $\mathcal{Q}_{d}^{n'}$ similar to (\ref{broad_mon}):
\begin{align}
\sum_{i=1}^{n-k+2}\mathcal{Q}_{d}^{n'}(A,B^i)\leq (n-k+2)(n+d).\label{broad_mon_q}
\end{align}

\subsection{New inequalities for Dicke states}
We now introduce a set of inequalities which can show strict monogamy of Dicke states in the high $n$ limit. These are based on  recent work by Heaney, Cabello, Santos and Vedral~\cite{1367-2630-13-5-053054} where they show that for the W state, it is possible to construct nonlocality tests and inequalities that are ``maximal'' in some sense, i.e. the violation of the inequality goes to the algebraic maximum in the $n\rightarrow\infty$ limit, thus mimicking perfect correlations of stabilizer states and the Mermin inequality~\cite{PhysRevA.77.062106}~\cite{PhysRevLett.65.1838}.
The inequality introduced in~\cite{1367-2630-13-5-053054} by Heaney, Cabello, Santos and Vedral (hereafter referred to as the HCSV inequality), has the property that the larger $n$ is, the higher the violation becomes. Although the original HCSV inequality only works for W states, it can be extended as follows to cover all Dicke states.

Following and extending the reasoning in~\cite{1367-2630-13-5-053054} for W state, if all $n$ parties measure in the $\sigma_z$ basis on a Dicke state $|S(n,k)\rangle$, $n-k$ of them will get result $0$ and the other $k$ will get result $1$ with certainty (though it is impossible to know who gets what). Now imagine that when $n-k-1$ parties get $0$ and the other $k-1$ parties get $1$, the remaining two decide instead to measure $\sigma_x$. In this case they will always get the same result. Since under LHV the results of one party should not depend on other parties' settings, this means that should any two chose to measure in $\sigma_x$, they would get the same result. If these results are given by an LHV distribution, this would mean that if all parties were to measure in $\sigma_x$ in the beginning, they should all get the same result. Since everything above occurs with certainty, we should always see, under LHV, that if all parties measure $\sigma_x$ they get the same result. However, simple calculation shows that this is not the case for all Dicke states.

The associated Bell inequality is
\begin{align}
\mathcal{L}= &\sum P(\pi(\underbrace{0\dots0}_{n-k}\underbrace{1\dots1}_{k})|0\ldots0) \nonumber\\
-&\sum P(\pi(\underbrace{0\dots0}_{n-k-1}\underbrace{1\dots1}_{k-1}01)|\pi(\underbrace{0\dots0}_{n-2}11))\nonumber\\
-&P(0\ldots0|1\ldots1)-P(1\ldots1|1\ldots1) \leq 0,\label{heaneyext}
\end{align}
where the permutations in the second and third lines are over parties fixing the relationship between measurement settings and results, as with $\mathcal{P}^n$. To see that this cannot be violated under LHV it is sufficient to see that it cannot be violated for any deterministic strategy (i.e. taking marginal probabilities to be zero or one)~\cite{PhysRevA.64.032112}, since all LHV distributions can be considered as probabilistic mixtures of deterministic ones. It is not difficult to see that taking any one of the $P(\pi(\underbrace{0\dots0}_{n-k}\underbrace{1\dots1}_{k})|0\ldots0)$ to be one cannot be compatible with keeping all the negative terms zero. Since these are the only possible positive terms, and at most only one can be equal to one, for all deterministic local strategies the expression is non-positive and a violation is incompatible with LHV. For a Dicke state $\ket{S(n,k)}$, $\mathcal{L}$ is violated by $1-\frac{{n\choose k}}{2^{n-1}}$. As for the $W$ state considered in~\cite{1367-2630-13-5-053054}, this achieves the algebraic maximum in the limit of large $n$, imitating perfect correlations of GHZ and other stabilizer states. This also implies strict monogamy for the limit in $n$. 

\begin{figure}[htbp]
\begin{center}
\includegraphics[width=230px,keepaspectratio=true]{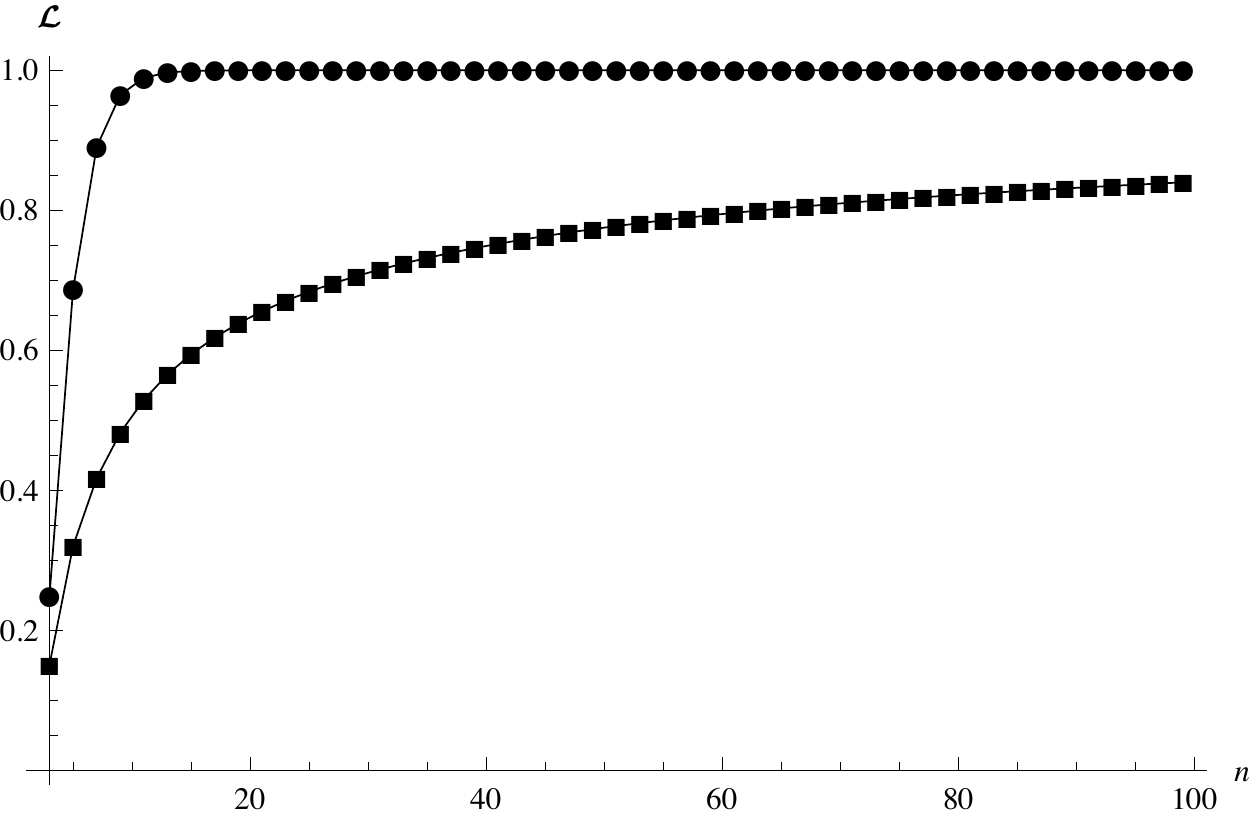}
\caption{A comparison of the violation of $\mathcal{L}$ (\ref{heaneyext}) for the states $\ket{S(n,\frac{n}{2})}$ (\ding{110}) and $\ket{W_n}$ (\ding{108}) as a function of $n$ (the number of parties).}
\label{heaney_plot}
\end{center}
\end{figure}

We plot the violation of $\mathcal{L}$ for $\ket{S(n,\frac{n}{2})}$ and $|W_n\rangle$ in Fig.~\ref{heaney_plot}. We see that the W state reaches one more quickly, in keeping with its lower entanglement.

\section{Conclusions and Discussions}

In this work we have studied the nonlocal properties of symmetric states as exposed by a set of inequalities. We have used the Majorana representation, numerics and semidefinite programming approaches to look at how classes of states can be identified using the inequalities, the scaling in $n$ for GHZ and W states and what we can say about monogamy of correlations that are seen.

Concerning types of entangled states, we have been able to separate LUP and LU classes of states for four qubits using our inequalities, hence in a device independent way. The example states chosen also sit in different SLOCC classes. This was done by bounding the possible violation of inequalities using SDP techniques. Going above four qubits seems difficult as the numerics quickly get difficult with more parties, though simple basis checking numerics indicate that the W and GHZ states may be separated in this way. This furthers the discussion about how entanglement classifications can be interpreted using nonlocal features. On the one hand we have the general statement that degeneracy of MPs guarantees persistency of correlations \cite{PhysRevLett.108.210407} to subsystems. This is true for all states, not just specific examples such as those expended upon here. We see that certain ``example" states such as the $|000+\rangle$ and $|W\rangle$ states may be separated from less degenerate states using this fact. This can be compared to the robustness of nonlocality under system loss~\cite{PhysRevLett.108.110501}~\cite{PhysRevA.86.042113}. On the other hand, we also saw an example with the GHZ and T states where $\mathcal{Q}^n_d$ can be used to discriminate different classes, not related to degeneracy ($|T\rangle$ and $|GHZ\rangle$, both with degeneracy one). Intriguingly, we also remark that these states naturally appear in the phase space of spinor condensates~\cite{PhysRevLett.99.190408}, pointing to a potential interest of these ideas in many-body physics, for example to witness different phases of matter where standard order parameters fail. Existing connections between entanglement classes and symmetry could further be useful in this direction~\cite{PhysRevA.83.042332}.

We also looked at what are the possible values violation can take. The first obvious statement with relation to entanglement was that the higher the entanglement is, the lower any possible violation of $\mathcal{P}^n$ and $\mathcal{Q}^n_d$ can be. At first this seems counterintuitive, but really it seems to stem from the simple fact that there is only one positive term - we later introduced larger inequalities with more positive terms based on the HCSV inequality~\cite{1367-2630-13-5-053054}, where the violation reaches its algebraic limit for all Dicke states in the high $n$ limit. We looked at how the violation of inequalities scale with $n$ for GHZ and W states. We see that W states fair much better for our inequalities, in contrast to the typical Mermin like inequalities where GHZ fairs better. We also look at the trends of the inequality violation with entanglement and see that this can be different. For W states and the $|000+\rangle$ state the violation increases with entanglement so that it gets closer to the upper bound ($\frac{1}{2^{E_g}}$), where as for GHZ states it goes down for higher $n$.

We then looked at what can be said about the monogamy of the correlations exposed by our inequalities and chosen measurement settings. First, we see that $\mathcal{P}^n$ and $\mathcal{Q}^n_d$ are not suited to showing strict monogamy (that is, we cannot say violation at the level achieved by quantum states implies no correlations are shared with another party), since, by the fact that entanglement bounds the violation, any quantum violation cannot reach the algebraic limit. This may indicate that these inequalities are not so useful for device independent security for example, although bounds on correlation sharing less than these strict ones may be of interest. To this end, using techniques from~\cite{PhysRevLett.102.030403} we bound how much correlations can be shared with the inequalities. We then define new inequalities based on the HCSV inequality, where we see that all Dicke states are strictly monogamous in the limit of high $n$, as has been seen before for W states~\cite{1367-2630-13-5-053054}. In this sense the extreme nonlocality of GHZ and stabilizer states seems to be replicated by Dicke states in the large $n$ limit. It remains open how general this is for all symmetric states.

One can also ask what other nonlocal properties can be inspected by inequalities $\mathcal{P}^n$ and $\mathcal{Q}^n_d$. Another property of multiparty correlations which is of interest, is whether it can be said to be ``genuine" or not - that is, whether the correlations at hand could be achieved by grouping the $n$ into subgroups or not. If not, we would say the correlations are genuinely $n$ party. The Svetlichny type inequalities~\cite{PhysRevD.35.3066} endeavor to identify this property - they should only be violated by genuinely $n$ party correlated states. Unfortunately it is not to hard to see that all the inequalities we use in this work do not have this property - it is possible to group parties together such that local states with respect to the new groupings can violate the inequalities. This can be easily seen by grouping the first $n-2$ parties and construct an LHV model by only using deterministic probabilities (probabilities equal to 0 or 1). The grouping makes it possible to set all negative terms to 0, and (one of) the positive term to 1. A stronger statement can be made by only grouping the first two terms - so that the weakest grouping still allows nonlocal correlations to violate all our inequalities. This is shown explicitly for $\mathcal{L}$ in the Appendix~\ref{proof}.

In summary it seems that one must make a balanced choice over which inequalities will be useful depending on circumstances. We have seen that $\mathcal{P}^n$ and $\mathcal{Q}^n_d$ are interesting in terms of separating classes of states, and indeed it is known to be true that all entangled pure states will show some violation $\mathcal{P}^n$~\cite{PhysRevLett.109.120402}. However, their violation can never be high enough to make the strongest statements we would like about monogamy. They also do not say whether correlations are ``genuine" or not (even $\mathcal{L}$, with its many positive terms, does not show genuine nonlocality or be maximally violated for finite $n$). On the other hand inequalities based only on expectation values (which necessarily have many positive terms) can have maximal violation for any $n$, but they cannot see the nonlocality of all states - there are entangled states which do not violate any inequality based on expectation values, which do violate $\mathcal{P}^n$~\cite{PhysRevLett.88.210402} . In a similar situation to the role of different entanglement measures in entanglement theory, it seems unlikely that any single inequality will be able to capture all the nonlocal properties we might be interested in.  

\begin{acknowledgments}
We thank Adel Sohbi for comments and discussions. We also thank Paul Jouguet for providing references regarding attacks on quantum key distribution systems. This work is supported by the joint ANR-NSERC grant ``Fundamental Research in Quantum Networks and Cryptography (FREQUENCY)''.
\end{acknowledgments}

\appendix
\section{Proof that $\mathcal{L}$ cannot detect genuine nonlocality}\label{proof}
To show that $\mathcal{L}$ cannot detect genuine nonlocality, we will group the first two parties and show that $\mathcal{L}=1$ under partially nonlocal LHV (where the first two parties are considered as one). Mathematically, an LHV model means that we can write
\begin{align}
P(a_1\ldots a_n|A_1\ldots A_n)=\int\! \rho(\lambda)\prod_{1\leq i\leq n} P_i(a_i|A_i,\lambda) \,\mathrm{d}\lambda,
\end{align}
where subscripts denote the parties.

Meanwhile, a partially nonlocal LHV means that we allow a subset of parties to be grouped together as a single (possibly nonlocal) party. In this proof, it means that
\begin{align}
&P(a_1\ldots a_n|A_1\ldots A_n)=\nonumber\\
\int\! \rho(\lambda)&P_{12}(a_1a_2|A_1A_2,\lambda)\prod_{3\leq i\leq n} P_i(a_i|A_i,\lambda)\,\mathrm{d}\lambda.
\end{align}

Below we give an explicit LHV model by setting all probabilities in $\mathcal{L}$ to equal to either 0 or 1. This implies only one term in the sum $\sum P(\pi(\underbrace{0\dots0}_{n-k}\underbrace{1\dots1}_{k})|0\ldots0)$ equals to $1$, all other terms will be $0$. Let us suppose, without loss of generality, 
\begin{align}
P(\underbrace{0\dots0}_{n-k}\underbrace{1\dots1}_{k}|0\ldots0)=1.
\end{align}

This implies
\begin{align}
&P_{12}(00|00)=1\\
&P_3(0|0)=1,\ldots,P_{n-k}(0|0)=1\\
&P_{n-k+1}(1|0)=1,\ldots,P_{n}(1|0)=1,
\end{align}
from which we can deduce
\begin{align}
&P_{12}(01|00)=P_{12}(10|00)=P_{12}(11|00)=0\label{00:a}\\
&P_3(1|0)=0,\ldots,P_{n-k}(1|0)=0\label{00:b}\\
&P_{n-k+1}(0|0)=0,\ldots,P_{n}(0|0)=0.\label{00:c}
\end{align}

For the terms $\sum P(\pi(\underbrace{0\dots0}_{n-k-1}\underbrace{1\dots1}_{k-1}01)|\pi(\underbrace{0\dots0}_{n-2}11))$, we will try to set all of them to $0$, using (\ref{00:a}) to (\ref{00:c}) with some extra probability assignments, without causing inconsistencies.

To see how we can set all terms to 0, first we divide the terms in the sum into three different cases ($a,b$ are both bits, $\bar{a}, \bar{b}$ denote their logical flip):
\begin{enumerate}
\item $P(ab\,\pi(0\ldots01\ldots01)|00\,\pi(0\ldots011))$.

In this case, if $a$ and $b$ are not both 0,then by (\ref{00:a}), the probability is 0. Otherwise, we can set $P_i(0|1)=0$, where $i\neq 1,2$.
\item $P(ab\,\pi(0\ldots01\ldots\bar{b})|01\,\pi(0\ldots001))$, $P(ab\,\pi(0\ldots01\ldots\bar{a})|10\,\pi(0\ldots001))$.

In this case, if $a=b=1$, then there exists $P_i(0|1)$ where $i\neq1,2$. Thus we can have $P_i(0|1)=0$ and $P_{12}(11|01)=1, P_{12}(11|10)=1$, without causing any inconsistency with the previous case. The latter two assignments also imply that if $a$ and $b$ are not both 1, then $P_{12}(ab|01)=P_{12}(ab|10)=0$.
\item $P(a\bar{a}\,\pi(0\ldots01\ldots01)|11\,\pi(0\ldots0))$.

In this case, the probability is always 0. This can be deduced from the pigeonhole principle: there are $n-k-1$ zero outcomes when parties $3$ to $n$ all measure in the 0 basis, so at least one party from $n-k+1$ to $n$ will get outcome 0 when measuring in the 0 basis. By (\ref{00:c}) the probability is 0.
\end{enumerate}

In the last case, because the probability is always 0 regardless of the probability assignments of the first two parties, we can set $P_{12}(00|11)=0$ and $P_{12}(11|11)=0$ without causing any inconsistency. These assignments guarantee that the last two probabilities in $\mathcal{L}$: $P(0\ldots0|1\ldots1)$ and $P(1\ldots1|1\ldots1)$, are 0.

Thus we can consistently assign probabilities such that all negative terms in $\mathcal{L}$ are 0 and the sum of all positive terms are 1, so $\mathcal{L}=1$, violating the inequality under partially nonlocal LHV. This shows that $\mathcal{L}$ cannot detect genuine nonlocality.
\qed

A similar argument can be made for $\mathcal{P}^n$ and $\mathcal{Q}^n_d$.

\bibliography{biblio}

\end{document}